\documentstyle[12pt]{article}
\begin{document}

\renewcommand{\thefootnote}{\fnsymbol{footnote} }

\begin{flushright}
BNL-63106
\end{flushright}

\begin{center}
{\bf An Exact Solution for Quantum Tunneling in a Dissipative System }\\%
\vspace{1cm}

Li Hua Yu\\

National Synchrotron Light Source, Brookhaven National Laboratory, N.Y.11973 %
\vspace{0.4cm}

Abstract\\
\end{center}

Applying a technique developed recently [1,2] for an harmonic oscillator
coupled to a bath of harmonic oscillators, we present an exact solution for
the tunneling problem in an Ohmic dissipative system with inverted harmonic
potential. The result shows that while the dissipation tends to suppress the
tunneling, the Brownian motion tends to enhance the tunneling. Whether the
tunneling rate increases or not would then depend on the initial conditions.
We give a specific formula to calculate the tunneling probability determined
by various parameters and the initial conditions.

\newpage\ 

\begin{center}
1. Introduction.
\end{center}

Quantum tunneling with dissipation has been studied by many people after the
work by Caldeira and Leggett[3-6]. These studies use different
approximations to calculate the tunneling probability. Among these works,
there is a widely discussed question about whether the dissipation
suppresses or enhances the quantum tunneling[7]. In this paper, we shall
show that for a special potential barrier, the inverted harmonic potential
well, the tunneling problem can be solved exactly, thus we shall answer this
question precisely. The technique we used for this solution has been
developed recently to solve the wavefunction evolution for another
dissipative system problem: an harmonic oscillator coupled to an
environment[1,2]. In this paper we shall show that this technique is equally
applied to the inverted harmonic potential.

The simplest example of a dissipative system, an harmonic oscillator coupled
to an environment of a bath of harmonic oscillators, has been the subject of
extensive studies (see [1] for references). In a recent paper[1], we
obtained a simple and exact solution for the wave function of the system
plus the bath, in the Ohmic case. It is described by the direct product in
two independent Hilbert spaces. One of them is described by the
Caldirola-Kanai (CK) Hamiltonian, the other represents the effect of the
bath, i.e., the Brownian motion. In a second paper[2], using this
wavefunction solution, we derived a simple formula for the probability
distribution at finite temperature, expressed in terms of the wavefunction
solution for the CK Hamiltonian only.

In this paper we shall apply these results to a different system, the
inverted harmonic potential well to study the quantum tunneling problem of
dissipative system. We shall use the same notations as the reference [1,2]
and quote formulas from there. The Hamiltonian of this system is:

\begin{equation}
H=\frac{p^2}{2M}-\frac 12M(\omega _0^2+\Delta \omega
^2)q^2+q\sum_jc_jx_j+\sum_j\left( \frac{p_j^2}{2m_j}+\frac 12m_j\omega
_j^2x_j^2\right) \;\;\;.
\end{equation}
The only difference of this Hamiltonian from that in reference [1] is the
sign of the second term of the right hand side, i.e., the potential is now
an inverted harmonic potential well. Assuming the Ohmic bath oscillator
density to be the same as given in [1]: 
\begin{equation}
\rho (\omega _j)=\frac{2\eta M}\pi \frac{m_j\omega _j^2}{c_j^2}.
\end{equation}
Following exactly the same procedure as [1], we derive the Langevin equation
of motion for the system:

\begin{equation}
\ddot q+\eta \dot q-\omega _0^2q=f(t)\,\,\,,
\end{equation}
with the Brownian motion driving force: 
\begin{equation}
f(t)=-\sum_j\frac{c_j}M(x_{j0}\,\cos \omega _jt+\dot x_{j0}\frac{\sin \omega
_jt}{\omega _j})\,\,\,.
\end{equation}

This equation is easily solved to give: 
\begin{equation}
q(t)=a_1(t)q_0+a_2(t)\dot q_0+\sum_j\,(b_{j1}(t)x_{j0}+b_{j2}(t)\dot
x_{j0})\,\,,
\end{equation}
where $q_0,\dot q_0,x_{j0},\dot x_{j0}$ are the initial position and
velocity operators of the system and bath respectively, and 
\begin{equation}  \label{a1a2}
a_1=e^{-\frac \eta 2t}(\cosh \omega t+\frac \eta {2\omega }\sinh \omega
t),\;a_2=e^{-\frac \eta 2t}\frac{\sinh \omega t}\omega ,
\end{equation}
\begin{equation}
b_{j1}\left( t\right) =-\frac{c_j}M\int_0^ta_2\left( t^{\prime }\right) \cos
\omega _j\left( t-t^{\prime }\right) dt^{\prime },\;b_{j1}\left( t\right) =-%
\frac{c_j}M\int_0^ta_2\left( t^{\prime }\right) \sin \omega _j\left(
t-t^{\prime }\right) dt^{\prime },
\end{equation}
with $\omega =(\omega _0^2+\frac{\eta ^2}4)^{1/2}$.

Then, based on the same arguments of [1], this solution can be used to show
that the wavefunction of the system plus the bath can be written as 
\begin{equation}
\Psi (q,\{\xi _j\},t)=\psi (q-\sum_j\xi _j,t)\,\prod_{j=1}^N\chi _j(\xi
_j,t)\,\,,  \label{wavefunction1}
\end{equation}
where the wavefunction $\psi (Q,t)$ is a solution of Schoedinger equation
with the time dependent CK Hamiltonian 
\begin{equation}
H_Q=e^{-\eta t}\frac{P^2}{2M}-\frac 12M\omega _0^2e^{\eta t}Q^2\,\,\,,
\label{hamiltonian}
\end{equation}
and the commutation $[Q,P]=i\hbar $, while $\xi _j=b_{j1}(t)x_{j0}+b_{j2}(t)%
\dot{x}_{j0}\,\,\,$is the contribution of the j'th bath oscillator to the
Brownian motion. The function $\chi _j(\xi _j,t)\,$ is given by $\chi _j(\xi
_j,t)\,\,\,\,\,\,\,\,\,=\,\,\,\,<\theta _{\xi _j}|\chi _{j0}>$, where $\chi
_{j0}(x_{j0})$ is the initial state of the bath oscillator, and $|\theta
_{\xi _j}>$ is an eigenstate of the operator $\xi _j\ $ with eigenvalue $\xi
_j$, as given by eq.(4) of [2]:

\begin{equation}
\theta _{\xi _j}(x_{j0},t)=\left( \frac{m_j}{2\pi \hbar b_{j2}}\right)
^{\frac 12}exp\left[ -\frac{im_j}{2\hbar b_{j2}}\left(
b_{j1}x_{j0}^2-2x_{j0}\xi _j\right) \right] .  \label{eqthetaj}
\end{equation}

According the analysis of [2], assuming the bath is at temperature T,
averaging over the Boltzmann distribution of the wavefunction $\chi
_{j0}(x_{j0})$ of the bath oscillators, and using eq.(\ref{wavefunction1}),
we find the probability distribution the same as derived in [2]:

\begin{equation}  \label{density1}
\rho (q,t)=\frac 1{\sqrt{2\pi }\sigma _\xi }\int |\psi (q-\xi ,t)|^2e^{-%
\frac{\xi ^2}{2{\sigma _\xi }^2}}d\xi ,
\end{equation}
where $\sigma _\xi $ is the Brownian motion width: 
\begin{equation}  \label{brownwidth1}
\sigma _\xi ^2(t)=\int\limits_0^{\omega _c}\frac \hbar {2m_j\omega
_j}(|b_{j1}(t)|^2+\omega _j^2|b_{j2}(t)|^2)\,\coth \left( \frac{{\ \hbar }%
\omega _j}{2kT}\right) \rho (\omega _j)d\omega _j.
\end{equation}

This result is apparently the same as that given in [2], but the expressions
for $b_{j1}(t)\ $and $b_{j2}(t)$ are different. There are also subtleties
associated with a logarithmic divergence of the integration over $\omega _j$%
, which is removed by introducing a cut-off frequency $\omega _c$. This
width is zero initially, but then increases exponentially to infinity as
time evolves. In the following, for simplicity, we consider only the low
temperature limit, then the width approaches in a time range of about 1/$%
\omega $ to the asymptotic value 
\begin{equation}
\sigma _\xi ^2=e^{(2\omega -\eta )t}\sigma _0^2\frac \eta {2\omega }\frac{%
\omega _0}\omega B,  
\label{brownwidth}
\end{equation}
with $\sigma _0^2\equiv \hbar /(2M\omega _0)\;$(the width of the ground
state of an harmonic oscillator with frequency $\omega _0$) and \ 
\begin{equation}
B\equiv \frac 1\pi \ln [1+(\frac{\omega _c}{\omega -\frac \eta 2})^2].
\label{B}
\end{equation}

\begin{center}
2. The wavefunction evolution
\end{center}

To calculate the tunneling probability, we assume that initially the
wavefunction is a gaussian wave packet centered at the right of the peak of
the potential by $z_0$ with a velocity $v_0=-\hbar k/M$ towards the
potential barrier peak:

\begin{equation}
\psi _0(q)\equiv \psi (q;t=0)=(2\pi \sigma ^2)^{-1/4}e^{\displaystyle -\frac{%
(q-z_0)^2}{4\sigma ^2}+ikq}.  \label{iniwave1}
\end{equation}
First, following exactly the same method in [1], we derive the Green's
function of the time dependent Hamiltonian eq.(\ref{hamiltonian}):

\begin{equation}
G(q,q_0;t,0)=\left( \frac M{2\pi i\hbar a_2}\right) ^{\frac 12}exp\left[ 
\frac{iM}{2\hbar a_2}\left( a_1q_0^2+\dot a_2e_1^{\eta t}q^2-2q_0q\right)
\right] {\it {\rm .}}  \label{green}
\end{equation}
Then we calculate the wavefunction $\psi (q,t):$

\begin{eqnarray}
\psi (q,t) &=&\int G(q,q_0;t,0)\psi _0(q_0)dq_0 \\
\ &=&(2\pi \sigma ^2)^{-1/4}\,(a_1+i\omega _0a_2r^2)\,\,e^{\displaystyle -%
\frac{(q-z_0)^2}{4\sigma ^2}+i(c_2q^2+c_1q+c_0)}  \label{wavefunction}
\end{eqnarray}

This result represents a gaussian distribution centered at $%
q_c=a_1z_0-a_2v_0 $ ( the classical trajectory of a particle initially at $%
z_0$ with velocity --$v_0$) and with a width of

\begin{equation}
{\sigma _\theta ^2}\equiv \sigma ^2({a}_1^2+r^4{\omega _0^2}a_2^2),
\label{sigmatheta}
\end{equation}
where $r\equiv \sigma _0/\sigma $. The coefficients in the phase factor
exponent are:

\begin{equation}
c_2=\frac{Me^{\eta t}}{4\hbar }\frac d{dt}(\ln \sigma _\theta ^2),
\label{c2}
\end{equation}
\begin{equation}
c_1=\frac{Me^{\eta t}}{4\hbar }(-2q_c\frac d{dt}(\ln \sigma _\theta
^2)+4\dot q_c),  \label{c1}
\end{equation}
\begin{equation}
c_0=\frac{ka_2}4e^{\eta t}(q_c\frac d{dt}(\ln \sigma _\theta ^2)-2\dot q_c)+%
\frac{q_cz_0}{4\sigma _\theta ^2}{\omega _0}a_2r^2.
\end{equation}
This phase factor is related to the current density, as will be explained
later. It is straight forward to verify that eq.(\ref{wavefunction}) indeed
satisfies the Schoedinger equation with the time dependent Hamiltonian eq.(%
\ref{hamiltonian}). With these provisions, we are ready to calculate the
tunneling probability.

\begin{center}
3. The tunneling probability and current density
\end{center}

Using the density eq.(\ref{density1}) and wavefunction eq.(\ref{wavefunction}%
), we find a very simple expression for the probability density: 
\begin{equation}
\rho (q,t)=\frac 1{\sqrt{2\pi }\sigma _t}e^{-\frac{(q-q_c)^2}{2\sigma _t^2}},
\end{equation}
where

\begin{equation}  \label{sigmat}
{\sigma _t^2}\equiv \sigma _\theta ^2+{\sigma }_\xi ^2
\end{equation}
is the total width including the Brownian motion width.

The probability for the particle to pass to the left of position q at t is
then:

\begin{equation}
P(q,t)=\int_{-\infty }^q\rho (q^{\prime },t)dt^{\prime }=F(\frac{q-q_c}{%
\sqrt{2}\sigma _t}),  \label{probability}
\end{equation}
where the function $F$ is:

\begin{equation}
F(W)\equiv \frac 1{\sqrt{\pi }}\int_{-\infty }^We^{-u^2}du.
\end{equation}
Hence the current density is found to be: 
\begin{equation}
I=-\frac{\partial P(q,t)}{\partial t}=|\psi |^2(\frac 12(q-q_c)\frac
d{dt}(\ln \sigma _t^2)+\dot q_c).  \label{i1}
\end{equation}

On the other hand, if the Brownian motion can be ignored (replacing $\sigma
_t$ by $\sigma _\theta $ ), the solution eq.(\ref{wavefunction}) of the time
dependent Schoedinger equation should also satisfy the expression for the
current density: 
\begin{equation}
I=\frac \hbar {2Me^{\eta t}}(\psi \frac{\partial \psi ^{*}}{\partial q}-\psi
^{*}\frac{\partial \psi }{\partial q})=\frac \hbar {Me^{\eta t}}|\psi
|^2(2c_2q+c_1).  \label{i2}
\end{equation}
The expressions for $c_1$ and $c_2$ eq.(\ref{c1}) and eq.(\ref{c2}) indeed
satisfy eq.(\ref{i1}), hence they acquire the following clear physical
meaning. If we denote the velocity of the ''fluid'' distribution $|\psi |^2$
as $v$, the current density is then $I=|\psi |^2$ $v$. Therefore a
comparison of eq.(\ref{i1}) with eq.(\ref{i2}) leads to:

\begin{equation}
v=\frac \hbar {Me^{\eta t}}(2c_2q+c_1)=\frac 12(q-q_c)\frac d{dt}(\ln \sigma
_\theta ^2)+\dot{q}_c.
\end{equation}
Thus those particles with larger q acquire larger velocity falling from the
apex of the potential. Since the width $\sigma _\theta ^2$ is proportional
to $\exp (2\omega -\eta )t$, as time evolves, most particles move rapidly
away from the apex of the potential. In the mean time, for any fixed $q\neq
0 $, the wavenumber $2c_2q+c_1$ also increases exponentially by a factor $%
e^{\eta t}$ according to eq.(\ref{c1}) and eq.(\ref{c2}) because the
equivalent mass is $Me^{\eta t}$ instead of $M$. Hence the wavelength decays
exponentially, as pointed out by reference [2]. When $t\gg 1/\eta $ the
system approaches the classical limit.

\begin{center}
4. Tunneling probability as time approaches infinity
\end{center}

We define the tunneling probability $P$ as the probability for the particle
to be at the left of the peak. Then, applying eq.(\ref{probability}), eq.(%
\ref{sigmat}) and eq.(\ref{sigmatheta}), we find $P=F(-W),$with

\begin{equation}
W=\frac{q_c}{\sqrt{2}\sigma _t}=\frac{a_1z_0-a_2v_0}{\sqrt{2(\sigma
^2(a_1^2+r^4{\omega _0^2}a_2^2)+\sigma _\xi ^2)}}.
\end{equation}

Because $a_1,a_{2,}$ and $\sigma _\xi $ are all proportional to $e^{(\omega
-\eta /2)t}$ as time approaches infinity, $W$ approaches a finite limit,
which determines the final tunneling probability. To simplify the expression
for this limit, we use the following scaled variables:

\begin{equation}
Z\equiv \frac{z_0}\sigma ;\;V\equiv \frac{v_0}{\omega _0z_0};\;\epsilon
\equiv \frac \eta {2\omega };\;r\equiv \frac{\sigma _0}\sigma ;\;B\equiv
\frac 1\pi \ln [1+(\frac{\omega _c}{\omega -\frac \eta 2})^2].
\end{equation}
Then, using the eq.(\ref{a1a2}) for $a_{1,}a_2$ , and eq.(\ref{brownwidth})
for the Brownian motion width, we have: 
\begin{equation}
W=\frac Z{\sqrt{2}}\frac{1-V\sqrt{\frac{1-\epsilon }{1+\epsilon }}}{\sqrt{1+%
\frac{1-\epsilon }{1+\epsilon }r^4+4B\epsilon \sqrt{1-\epsilon ^2}r^2}}.
\label{W}
\end{equation}

To understand the meaning of this expression, we remark that $Z\gg 1$
represents a case where the wavepacket width is much smaller than its
distance from the origin. We also remark that if $\eta =0$, as time
approaches infinity, $q_c=e^{\omega _0t}(z_0-v_0/\omega _0)/2$ . Hence $V<1$
means that a classical particle with initial velocity $v_0$ and position $%
z_0 $ does not have enough kinetic energy to pass the potential barrier if
there is no dissipation. Thus $Z\gg 1$ and $V<1$ represent a case relevant
to the quantum tunneling problem.

The term $4B\epsilon \sqrt{1-\epsilon ^2}r^2$ comes from the Brownian
motion. If $\epsilon \ll 1$, i.e., if the dissipation is small, this term
increases as $\epsilon $ increases, and reduces $W$, which in turn increases 
$F(-W)$, hence enhances the tunneling. This effect is reduced if $r\ll 1$,
or, in other words, if the initial wavepacket width is much larger than $%
\sigma _{0}$the increase due to The Brownian motion is
insignificant. If the initial velocity is not zero, the second term in the
numerator increase $W$ as $\epsilon $ increases, thus the damping suppresses
the tunneling. Intuitively this can be explained as that the damping makes
the classical particle unable to move to the barrier peak as close as if
there is no damping, and hence increases the barrier height.

To get an idea about the effect of dissipation on the tunneling probability,
in figure 1, we plot $P$ as a function of $V$ and $\epsilon $ for $Z=3$, $%
r=0.3$, and $B=3$, which corresponds to a cut-off frequency about 100 time
larger than $\omega $. Because the logarithmic dependence of $B$ on $\omega
_c$, the result is very insensitive to the cut-off frequency. The plot
clearly shows that when $V$ is large, the tunneling is suppressed by the
dissipation, while if initial velocity is zero, the dissipation enhances the
tunneling.

\begin{center}
5. Tunneling probability and uncertainty principle
\end{center}

Finally, it is interesting to examine the relation between the initial
momentum and position distribution and the tunneling probability. For
simplicity, we consider here only the case without dissipation. The Fourier
transform of initial wavefunction eq.(\ref{iniwave1}), the wavefunction in
momentum space is:

\begin{equation}
\psi _0(k)=(\frac{2\sigma ^2}\pi )^{1/4}e^{-(k+k_0)^2-i(k+k_0)z_0}.
\end{equation}

Thus from the classical point of view, the probability for the particle to
pass the potential barrier is the probability of the initial velocity $%
v<-\omega _0z_0$, or $k<-z_0/2\sigma _0^2$ (notice that $v=\hbar k/M$):

\begin{equation}
P=\int_{-\infty }^{-\frac{z_0}{2\sigma _0^2}}|\psi _0(k)|^2dk=F(-\frac{Z(1-V)%
}{\sqrt{2}r^2}).
\end{equation}
A comparison with eq.(\ref{W}) shows that if $r\gg 1$ , i.e., if the initial
width $\sigma $ is much smaller than the minimum wavepacket width $\sigma _0$%
, or, in other words, if the initial momentum spread is very large, this
crude estimate is correct.

Another extreme is when $r\ll 1$, i.e., when we can consider the initial
momentum spread is very small, and all the particles have velocity $v_0$,
but initial position has large spread. Then from the classical point of
view, the probability for particle to pass the potential barrier is its
initial $q<v_0/\omega _0$, i.e.:

\begin{equation}
P=\int_{-\infty }^{v_0/\omega _0}|\psi _0(q)|^2dq=F(-\frac{Z(1-V)}{\sqrt{2}}%
).
\end{equation}
Again, a comparison with eq.(\ref{W}) shows that if $r\ll 1$ , i.e., if the
initial width $\sigma $ is much larger than the minimum wavepacket width $%
\sigma _0$, this crude estimate is also correct. These estimates indicate
that the tunneling probability has a very simple relation with the
uncertainty principle, and we have a very simple way to estimate the
tunneling probability.

\begin{center}
{\bf Acknowledgments}
\end{center}

The author thanks Prof. C.N. Yang for many sessions of stimulating
discussions. The author also likes to thank C.P. Sun for interesting
discussions. The work is performed under the auspices of the U.S. Department
of Energy under Contract No. DE-AC02-76CH00016.

\newpage

\begin{center}
{\bf References}
\end{center}

\vspace{0.5cm}

\begin{enumerate}
\item  L.H.Yu, C.P.Sun, Phys. Rev. A {\bf 49}, 592 (1994)

\item  L.H.Yu, Phys. Lett. A {\bf 202}, 167, (1995)

\item  A.O. Caldeira, A.J. Leggett, Phys. Rev. Lett. {\bf 46, }211{\bf \ (}%
1981); Ann. Phys. {\bf 149}, 374 (1983)

\item  A.J.Bray and M.A.Moore, Phys. Rev. Lett. {\bf 46, }1545{\bf \ (}%
1982); S. Chakravarty, ibid. 49, 681 (1982);

\item  G. Sch\"{o}n and A.D.Zaikin, phy. Rep. 198, 237 (1990), and
references therein.

\item  S. Chakravarty and A.J. Legget, Phys. Rev. Lett. 52, 5 (1984);
A.J.Leggett, S. Chakravarty, A.T. Dorsey, M.P.A.Fisher, A.Garg, and W.
Zweger, Rev. Mod.Phys. 59,1 (1987).

\item  About the discussion on whether the dissipation suppresses or enhance
the quantum tunneling, see for example: A. J. Leggett; Satoshi Iso; in
''Proceedings of the 4th International Symposium: Foundation of Quantum
Mechanics in the Light of New Technology''; K. Fujikawa, S. Iso, M. Sasaki,
and H. Suzuki, Phys. Rev. Lett. {\bf 68, }1093{\bf \ (}1992).

\vspace{0.5cm}
\end{enumerate}

\newpage

\begin{center}
{\bf Figure Captions}
\end{center}

Figure 1. The tunneling probability as a function of the scaled dissipation
coefficient $\epsilon $ and the scaled initial velocity $V$, assuming the
scaled initial position $Z=3,$ the scaled inverse wavepacket size $r=0.3,$
and the scaled cut-off frequency $B=3.$

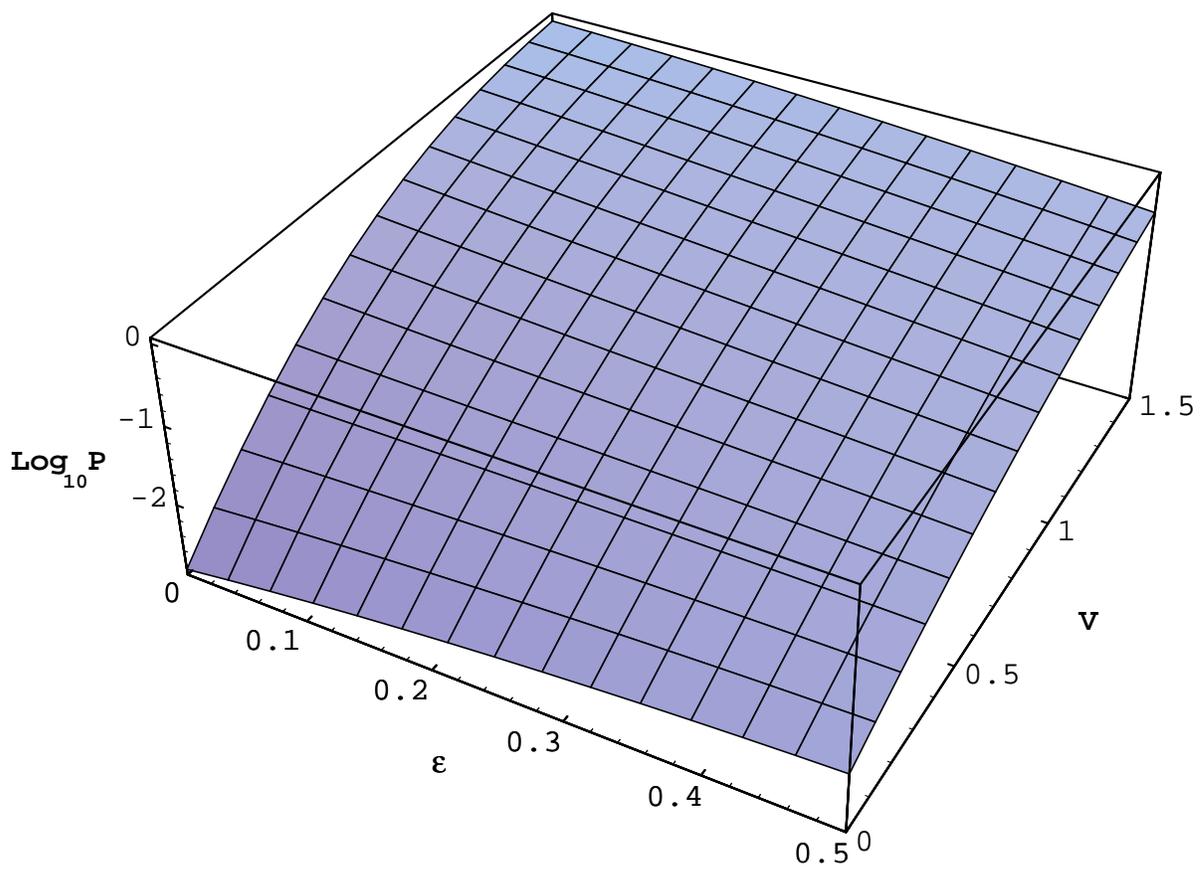
\end{document}